%% file: main.tex
\documentclass[journal]{IEEEtran}

\usepackage{cite}

\usepackage{hyperref}
\usepackage{xcolor}
\usepackage{booktabs}
\usepackage{multirow}
\usepackage{tablefootnote}
\usepackage{array}
\usepackage{authblk}
\usepackage{xspace}

\ifCLASSINFOpdf
  \usepackage[pdftex]{graphicx}
\else
\fi
\usepackage{subcaption}

\usepackage{amsfonts, amsmath, amssymb}
\usepackage{pifont}
\usepackage{amsmath}
\usepackage{bbold}

\usepackage[]{mdframed}
\usepackage{algpseudocode}

\usepackage{url}

\newcommand{\titleheader}[1]{\vspace{0.3cm}\small\textit{#1}\vspace{0.3cm}}


\begin{document}
%
\title{BlackCATT: Black-box Collusion Aware Traitor Tracing in Federated Learning}

%
%
%
\author[1]{Elena Rodríguez-Lois,~\IEEEmembership{Graduate Student Member,~IEEE}
\thanks{Work partially funded by the EU (Next Generation) through the "Plan de Recuperación, Transformación y Resiliencia" under project FELDSPAR: "Federated Learning with Model Ownership Protection and Privacy Armoring" (Grant MCIN/AEI/10.13039/501100011033), by the Spanish Ministry of Science, Innovation and Universities via a doctoral grant to the first author (FPU22/01929), and by Xunta de Galicia and the European Regional Development Fund, under project ED431C 2025/41.}}
\author[2]{Fabio Brau,~\IEEEmembership{Member,~IEEE}}
\author[2]{Maura Pintor,~\IEEEmembership{Member,~IEEE}}
\author[2]{Battista Biggio,~\IEEEmembership{Fellow,~IEEE}}
\author[1]{Fernando Pérez-González,~\IEEEmembership{Fellow,~IEEE}}

\affil[1]{Signal Theory and Communications Department, atlanTTic Research Center, University of Vigo, Spain}

\affil[2]{Department of Electrical and Electronic Engineering, University of Cagliari, Italy}

\renewcommand{\subsectionautorefname}{Sect.}
\renewcommand{\sectionautorefname}{Sect.}
\renewcommand{\figureautorefname}{Fig.}


\maketitle

\titleheader{This work has been submitted to IEEE Transactions on Information Forensics and Security (TIFS) for possible publication.}


%
\IEEEpeerreviewmaketitle

\begin{abstract}
Federated Learning has been popularized in recent years for applications involving personal or sensitive data, as it allows the collaborative training of machine learning models through local updates at the data-owners’ premises, which does not require the sharing of the data itself. 
Considering the risk of leakage or misuse by any of the data-owners, many works attempt to protect their copyright, or even trace the origin of a potential leak through unique watermarks identifying each participant's model copy. 
Realistic accusation scenarios impose a black-box setting, where watermarks are typically embedded as a set of sample-label pairs.
The threat of collusion, however, where multiple bad actors conspire together to produce an untraceable model, has been rarely addressed, and previous works have been limited to shallow networks and near-linearly separable main tasks. 
To the best of our knowledge, this work is the first to present a general collusion-resistant embedding method for black-box traitor tracing in Federated Learning: BlackCATT, which introduces a novel collusion-aware embedding loss term and, instead of using a fixed trigger set, iteratively optimizes the triggers to aid convergence and traitor tracing performance.
Experimental results confirm the efficacy of the proposed scheme across different architectures and datasets. Furthermore, for models that would otherwise suffer from update incompatibility on the main task after learning different watermarks (e.g., architectures including batch normalization layers), our proposed BlackCATT+FR incorporates functional regularization through a set of auxiliary examples at the aggregator, promoting a shared feature space among model copies without compromising traitor tracing performance.   

\end{abstract}

\section{Introduction}

Deep Neural Networks (DNNs) have consistently proven their potential in recent years, long surpassing human performance across numerous applications~\cite{He15}, and fueling the growth of artificial intelligence into a multi-billion-dollar industry~\cite{newsMarket}. With its widespread adoption and monetization, discussions around copyright are becoming increasingly common, both concerning the models themselves, which could be unlawfully exploited by competitors~\cite{newsOpenAI}, and the training data~\cite{newsACTU}, which remains the basis of DNNs' success. In the latter case, the issue extends beyond economics, as collecting large amounts of examples is often unfeasible or unsafe for tasks involving personal or sensitive information. From this limitation emerged Federated Learning (FL)~\cite{McMahan16}, a decentralized training approach that allows multiple data-owners (or clients) to collaboratively train a model, sharing updates derived from training locally on their private data, without explicitly exposing the data itself. These FL collaborations, often made up of altruistic data-owners, further raise critical questions about ownership, accountability, and rightful use of the trained model.

Following previous works on DNN watermarking~\cite{Li21}, there have been numerous efforts over recent years to protect the copyright of FL models specifically, and the data-owners' claims to their contributions~\cite{Lansari23}. Furthermore, the need for trusted FL systems naturally aligns with the problem of traitor tracing in traditional watermarking theory~\cite{Chor94}, where the goal is not only to prove the FL coalition's ownership of a suspicious model, but also to trace any leak back to the malicious participants involved, enabling further action to hold them accountable within the system. This can be achieved through the use of unique watermarks\footnote{Traditionally referred to as fingerprints, though in the context of proving ownership of DNNs, fingerprints are typically understood as inherent properties of the trained models, not deliberately embedded watermarks.} identifying different model copies accessible to specific data owners, as explained in Section~\ref{sec:ttinfl}.

Most traitor tracing works in FL embed unique watermarks directly into the numerical values or internal features of the model parameters~\cite{Fang-Qi_Li21,Liang23,Shao23,Yu23,Luo24,Chen24_2,RodriguezLois24,Chen25}, requiring full access to a suspicious model copy to trace its origin (white-box). While this approach presents notable strengths, such as a higher watermarking capacity and reliability, it is often impractical, as it is unlikely that the suspicious model weights will be open to public scrutiny. To circumvent this limitation, some of these works also embed a system watermark into the input-output behavior of all model copies (black-box)~\cite{Shao23,Chen24_2} through specific examples known as triggers, akin to a backdoor attack~\cite{BadNets}, hoping that sufficient proof of ownership can be gathered to eventually grant white-box inspection of the model. However, such access cannot be assumed in practice, and this can lead to distrust and halting of collaboration among the participants. The use of unique black-box watermarks, which would allow the FL system to trace the leak directly once a suspicious model is available online, has received considerably less attention thus far~\cite{Fang-Qi_Li21,RodriguezLois24,RodriguezLois25,Xu25}. While a fully black-box approach would be more practical, black-box embedding can negatively impact the main task to a greater extent, making the training considerably more challenging than in the white-box setting~\cite{RodriguezLois24}. This practical difficulty is likely a key reason why unique black box watermarking has remained underexplored in the literature. 

Additionally, two or more malicious participants might collaborate to steal their models and create a new model (e.g., by averaging the weights of their own) that might result untraceable. 
In fact, while the effect of collusion could be straightforward to assess for a white-box scheme, its impact on the learned backdoors of a ``merged'' model is difficult to predict, due to the complexity and non-linearity of neural networks. 
This type of attack, known as \textit{collusion}, has been extensively studied in the traitor tracing literature~\cite{Boneh98}. 
However, this threat is largely overlooked in FL, with some exceptions~\cite{Luo24,RodriguezLois24,RodriguezLois25}. 

To the best of our knowledge, the work in~\cite{RodriguezLois23} is the only existing approach to attempt black-box traitor tracing on classifiers that considers collusion as a potential attack, later extended to FL models in~\cite{RodriguezLois24}. Because of the challenges of modeling the collusion on the backdooring examples,~\cite{RodriguezLois23} uses collusion-agnostic traitor tracing codes, namely $q$-ary Tardos codes~\cite{Skoric12}, that take advantage of the full dimensionality of the output and do not rely on a specific collusion operation to work. Also, in contrast to other unique backdoor-based watermarks~\cite{Fang-Qi_Li21,Xu25,Yu23}, the original work in~\cite{RodriguezLois23} argued in favor of a shared trigger set with different label vectors for all model copies, instead of non-overlapping trigger-label pairs. This new configuration allows a significant reduction in the number of queries needed to identify the source of a leak. However, these previous works~\cite{RodriguezLois23,RodriguezLois24,RodriguezLois25} remain proof of concept studies for Tardos-based black-box traitor tracing, described in Section~\ref{sec:ttinfl:tt}. Their results are limited to a shallow DNN and a simple near-linearly separable main task (MNIST~\cite{MNIST}), and they fail to provide a suitable embedding method that generalizes across more complex architectures and datasets. In such settings, the increased non-linearity of the learning function makes the result of a collusion attack hard to predict: the triggers' outputs often become largely uncorrelated with those of the original models, rendering effective tracing virtually impossible. 

In this work we propose BlackCATT, the first practical \textbf{Black}-box \textbf{C}ollusion-\textbf{A}ware \textbf{T}raitor \textbf{T}racing scheme that generalizes across architectures and is suitable for FL systems, explained in Section~\ref{sec:catt}. Our main contributions are:

\begin{enumerate}
    \item Highlighting that standard watermark embedding for traitor tracing is inherently vulnerable to collusion attacks.
    \item Introducing a collusion-aware loss term that explicitly optimizes the embedding to promote collusion-resistant behavior.
    \item Demonstrating that trigger set optimization through adversarial perturbations further enhances the traitor tracing capabilities of the scheme, enabling a more efficient detection even in early stages of training.
    \item Proposing functional regularization to prevent model drift between the different model copies, protecting the performance on the shared main task.
\end{enumerate}

Experimental results presented in Section~\ref{sec:exp} show the effectiveness and robustness of this approach, both against collusions of multiple data-owners and against further attacks, such as finetuning and pruning. This paper focuses on the specific challenge of black-box traitor tracing in FL, and given its limited scope, Section~\ref{sec:relworks} highlights other relevant research areas and works, while Section~\ref{sec:concl} discusses the main limitations, conclusions of the study, and future work that could improve the proposed scheme. To support reproducibility, we release our implementation at \url{https://github.com/erodriguezlois/BlackCATT}.

{\em Notation:} Lower-case bold letters (e.g. $\mathbf{a}$) represent column vectors, where their $i$th element is denoted with a subindex (e.g. $a_i$). Bold upper-case letters (e.g. $\mathbf{B}$) represent two-dimensional matrices. Calligraphic fonts (e.g. $\mathcal{C}$) represent sets, alphabets or bases. The function of the model with parameters $\boldsymbol{\theta}$ is represented as $f_{{\boldsymbol{\theta}}}(\cdot)$, with $f_{{\boldsymbol{\theta}}}(\mathbf{a}) = \mathbf{b}$ the soft output vector to input $\mathbf{a}$.

\section{Traitor Tracing in Federated Learning} \label{sec:ttinfl}

\subsection{Federated Learning Definitions and Threat Model} \label{sec:ttinfl:fl}

In FL systems, multiple data-owners collaborate to train one or several models, typically because they lack sufficient data to train one independently. To coordinate and aggregate the local updates of each data-owner, FL systems usually rely on a central node referred to as the aggregator, which does not receive data but only model updates from the data owners. A standard FL scheme proceeds as follows: 

\begin{enumerate}
    \item Let there be $N$ data-owners in the FL system, each identified by an integer index $j$, and with $\mathcal{N}$ representing the set of all their indexes. The aggregator initializes several model copies $\boldsymbol{\theta}_j^{(0)} \forall j \in \mathcal{N}$ with the same parameters. 
    \item For each round $r$:
    \begin{enumerate}
    \item A uniformly sampled random subset $\mathcal{P}^{(r)}\subseteq \mathcal{N}$ of $P$ data-owners is called to participate, and train their current model $\boldsymbol{\theta}_j^{(r)}$ on their private data to obtain a locally updated version $\hat{\boldsymbol{\theta}}_j^{(r)}$.
    \item The aggregator retrieves these locally trained versions $\hat{\boldsymbol{\theta}}_j^{(r)}$ $\forall j\in \mathcal{P}^{(r)}$, generates model copies for the next round $\boldsymbol{\theta}_j^{(r+1)}$ $\forall j\in \mathcal{N}$, and sends these back to each data-owner.\footnote{See Section \ref{sec:catt} for details into this step.}
    \end{enumerate}
\end{enumerate}

With this in mind, the following summarizes the roles and assumptions of the different participants in this work, as well as the threat model considered. 

\paragraph{Aggregator} We assume an honest and trusted aggregator that does not have direct knowledge of the main task (e.g., no labeled dataset), but has white-box access to all model copies in the system. The aggregator is in charge of updating these copies with every training round and distributing them back to the data-owners. Through this update the aggregator can insert a unique watermark into the different model copies. If the model is stolen, either the aggregator or an independent verifier can attempt to detect these watermarks to identify the data-owners behind the leak.
\paragraph{Data-owners} We assume all data-owners, including malicious participants, ultimately aim to obtain a model that achieves great performance on the private data. Because of this, whenever they are called to participate in a training round, data-owners will train their model copy on their local dataset using the hyperparameters specified by the system, and will not attempt any kind of attack through their shared updates that could hinder the training process (e.g., poisoning attacks~\cite{biggio2012poisoning,Lyu23}). 

\begin{figure}[ht]
    \centering   \includegraphics[width=\columnwidth]{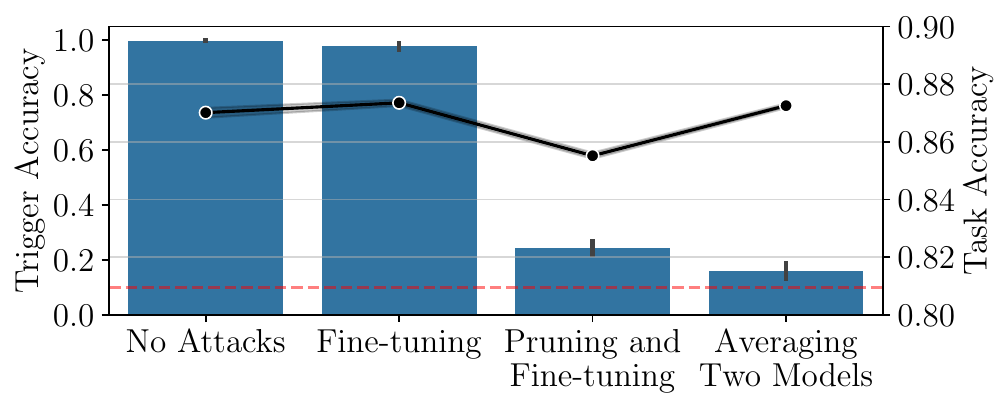}
    \caption{Accuracy of unique trigger-label sets (bar plot) and task accuracy (solid line) under different attacks, with a red dashed line indicating the random classification threshold for the trigger-label sets.}
    \label{fig:collusion_comparison}
\end{figure}

\textit{Threat of illegal exploitation and the overlooked attack}: At any point of the collaborative training process, if the model achieves reliable performance on their own task, data-owners could selfishly steal the model and use it for their own purposes. If one or multiple participants attempt to illegally steal the shared model and profit from it outside of the system agreement and without the consent of their collaborators, they may perform offline model edits with the intention to remove the traceable watermark, typically defined as a mapping $\mathcal{W} = \{\mathbf{X},\mathbf{y}\}$ between trigger examples and their assigned labels, memorized by the model. While testing individual model modifications (e.g., finetuning, pruning) is standard practice across all previous works, a collusion targeting the uniqueness of their individual watermark (e.g., parameter averaging, random sampling with other data-owners) can gravely affect the tracing capabilities of the schemes. This is evident in the comparison of Figure~\ref{fig:collusion_comparison}, which presents the accuracy of unique trigger-label pairs $\mathcal{W}_j = \{\mathbf{X}_j,\mathbf{y}_j\}$ embedded in the different model copies of each data-owner $j$, and demonstrates how averaging two models maintains high task accuracy while drastically reducing the effectiveness of the watermark's triggers.\footnote{Results from 20 data-owners performing FL with VGG16~\cite{VGG} and CIFAR-10~\cite{CIFAR10}, naively embedding different trigger-label pairs into each model copy.} Moreover, a collaborative attack as described above does not require any training data, is not computationally expensive, and does not have a detrimental effect on the model's performance. Given that a malicious participant could even pose as two different data-owners to gain access to multiple model copies, this threat is particularly relevant in the FL scenario. 

\begin{figure*}
    \centering
    \includegraphics[width=1\linewidth]{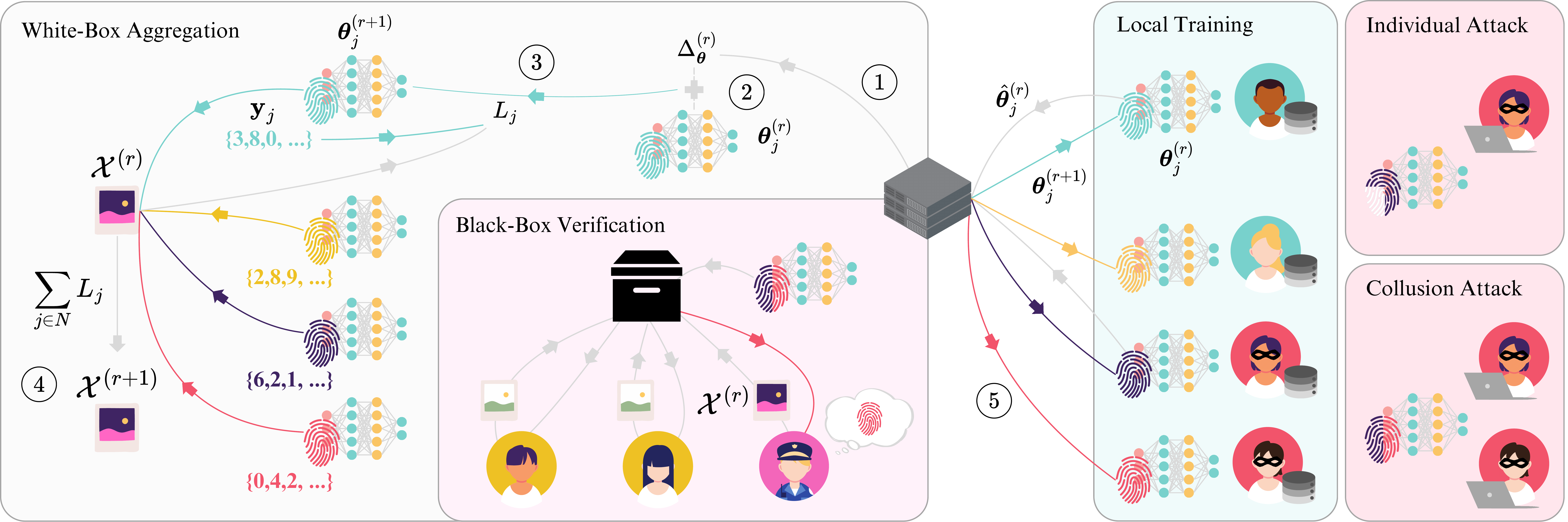}
    \caption{Conceptual representation of BlackCATT. The aggregator-side framework consists of five main steps: (1) \textit{Collection} of trained model copies; (2) \textit{Task Arithmetic} to update model copies; (3) \textit{Collusion-aware embedding} of the unique watermarks; (4) \textit{Optimization of the Shared Trigger Set} and (5) \textit{Return} to data-owners. For a leak involving one or multiple malicious participants, the watermarks can be verified through black-box queries, either through the aggregator or an independent verifier.}
    \label{fig:diagram_FL}
\end{figure*}

\subsection{Watermark and Traitor Tracing Requirements in FL}
The unique watermark embedded by the aggregator in FL to trace the identity of any malicious data-owner unlawfully exploiting their trained model should satisfy the standard requirements reported in previous works~\cite{Li21}:  \textit{Robustness} and \textit{Security} to resist both innocent processing (e.g., fine-tuning) and deliberate attacks (e.g., collusion), \textit{Fidelity} to not significantly alter the performance on the main task, \textit{Capacity} to embed enough identity information into the model copies, \textit{Integrity} to provide low false alarm and missed detection rates, \textit{Generality} to be applied to different architectures and datasets, and \textit{Efficiency} on the required computational overhead. In addition, because of the collaborative nature of the training in FL, the watermark must satisfy \textit{Fast Convergence} relative to the main task, ensuring that any intermediate model with acceptable performance throughout training is reliably watermarked.\footnote{This last requirement is often overlooked in previous works, e.g.~\cite{Xu25}, where the watermark embedding is not performed until the main task has sufficiently converged.}

\subsection{Accusation Process for Black-box Traitor Tracing} \label{sec:ttinfl:tt}
Let $\mathcal{X} = \{\mathbf{x}_i\}_{i=0}^{T-1}$ be a shared set of $T$ triggers, specific inputs used to verify any model copy through a unique watermark $\mathcal{W}_j = \{\mathbf{X},\mathbf{y}_j\}$ assigned to each data-owner, with $\mathbf{X} = [\mathbf{x}_1, \mathbf{x}_2, ..., \mathbf{x}_T]$, and $\mathbf{x}_i$ representing the $i$th trigger. According to $\mathcal{W}_j$, the output of each model copy to the trigger $\mathbf{x}_i$ is a label $y_{ji} \in \mathcal{Q}$, where $\mathcal{Q}$ is the set of all possible $q$ output classes in the classification task. When constructing $\mathcal{W}_j$, the label $y_{ji}$ is randomly sampled according to a secret $q$-component bias vector $\mathbf{p}_{i}$, that sets the probability of choosing each possible class $\ell\in\mathcal{Q}$ as $P[y_{ji}=\ell] = p_i^{(\ell)}$. This bias vector is drawn from a symmetric Dirichlet distribution, with $\kappa>0$ and a cutoff parameter $\tau$, so that $p_i^{(\ell)} \in [\tau, 1-(q-1)\tau] \forall i, \ell$~\cite{Skoric12}. By design, the bias vector $\mathbf{p}_i$ allows the outputs to some queries to be a more statistically significant evidence against certain users, according to how popular each label $\ell$ is among all data-owners. When encountering a suspicious unknown model, an iterative accusation process can start through the trusted aggregator or an independent verifier, where the predicted output label $\hat{y}_i = \text{argmax}(f_{{\boldsymbol{\theta}}}(\mathbf{x}_i) )$ to each query from the shared trigger set $\mathcal{X}$ incrementally adjusts a cumulative suspicious score for each data-owner $S_j^{(t)}$. This same cumulative approach is used in other Tardos-based works~\cite{Skoric12}. After $t$ queries, one would have
\begin{equation} \label{eq:cumscore}
    S_j^{(t)}=\sum_{i=1}^t S_{ji},
\end{equation}
where $S_{ji}$, is the suspicious score of data-owner $j$ for the query of $\mathbf{x}_i$. This suspicious score $S_{ji}$ will either increase the suspicion if the label matches the one assigned to the data-owner, or decrease it if it does not, considering the secret bias $p_i^{(y_i)}$ as
\begin{equation} \label{eq:score}
        S_{ji}= 
        \begin{cases}
        \sqrt{(1-p_i^{(y_i)})/p_i^{(y_i)}} & \text{if } y_{ji} = \hat{y}_i \\
        - \sqrt{p_i^{(y_i)} /(1-p_i^{(y_i)})} & \text{if } y_{ji} \neq \hat{y}_i
        \end{cases} .
\end{equation}
Following the theoretical analysis in~\cite{Skoric12}, a dynamic accusation threshold $Z^{(t)}$ can be set such that after $t$ queries, any data-owner whose cumulative score $S_j^{(t)}$ exceeds this value can be accused with a false positive probability not greater than $\epsilon_{\text{FP}}$:
\begin{equation} \label{eq:quad}
    Z^{(t)}=(\ln{\epsilon_{\text{FP}}})\cdot\left(-\frac{1}{3\cdot \sqrt{\tau}}\pm\sqrt{\frac{1}{9\cdot \tau}-\frac{2t}{\ln{\epsilon_{\text{FP}}}}}\right).
\end{equation}

This collusion-agnostic analysis provides guarantees on the permissible false positive rate.\footnote{Provided that the triggers and assigned labels are independent.} Tardos codes were originally designed as traitor tracing, collusion-agnostic codes, assuming that for a collusion of $c$ data-owners, defined as the set of their indexes $\mathcal{C}^c$ (e.g., for $c=3$, the collusion set would be $\mathcal{C}^3 = \{j_0,j_1,j_2\}$, representing the three colluding data-owners), any combination of their assigned codes will still be functional to identify at least one of the guilty participants through the overall suspicious scores $S_j$. This is because they rely on the \textit{Marking Assumption}~\cite{Skoric12}, which states that the expected output $\hat{y}_i$ of the colluded model to a certain query must be among the labels assigned to the colluders, $\mathcal{Y}_{\mathcal{C}i}$. While this assumption is very useful in the coding layer, it does not always hold for backdoor watermarking, as imperfect embedding, nonlinearities, and attacks will lead to misclassifications and unexpected behaviors. Because of this, the Marking Assumption Violation (MAV) rate, defined as
\begin{equation}\label{eq:mav}
    \text{MAV} = \frac{1}{T} \sum_{i=1}^{T} [{\hat{y}}_i \notin \mathcal{Y}_{\mathcal{C}i}],
\end{equation}
with $[\cdot]$ representing the Iverson bracket, serves as a measure of the trigger set's effectiveness for traitor tracing. A higher MAV, analogous to lower accuracy in the case of non-overlapping trigger-label pairs, means that many queries fail to provide useful evidence for the accusation, increasing the likelihood of missed detections and thus, a higher false negative rate.

\section{Collusion-Aware Traitor Tracing} \label{sec:catt}

An overview of the proposed framework BlackCATT is shown in Figure~\ref{fig:diagram_FL}, depicting the interaction between data-owners and the aggregator mentioned in Section~\ref{sec:ttinfl:fl}, as well as the responsibilities of the aggregator: updating the data-owners' model copies with the main and watermarking tasks, and optimizing the shared trigger set so that, in case of a leak, the black box watermark can be verified as described in Section~\ref{sec:ttinfl:tt}. 

In standard FL, the aggregator typically averages all updates into a single global model after each training round. When implementing aggregator-side watermarking, white-box schemes can typically re-embed unique watermarks from scratch from this global model, as the over-parametrization of DNNs and the spread spectrum nature of white-box schemes allows for many different ways to embed the watermark without significantly impacting the model's performance on the main task. On the other hand, black-box schemes need to modify the model's actual behavior. Because of this, a naive re-embedding of a black-box watermark is often either too disruptive, causing the main task to be forgotten, or not strong enough for detection, if the watermark is not actually memorized. Considering this, \textit{task arithmetic}~\cite{Ilharco23} is a useful alternative, aggregating only the parameter difference caused by the local training on the main task, and applying this aggregated main task vector into all watermarked model copies without erasing previous embedding efforts. Additionally, BlackCATT considers a novel collusion-aware embedding metric, and instead of static black-box triggers as done in prior work~\cite{Fang-Qi_Li21,Xu25,RodriguezLois24}, optimizes the shared trigger set to improve convergence and traitor tracing performance.

In the proposed approach each training round performed by the aggregator includes, as depicted in Figure~\ref{fig:diagram_FL}: 
\begin{enumerate}
    \item \textit{Collection}: Collecting each of the updated weights $\hat{\boldsymbol{\theta}}_j^{(r)}$ from $\mathcal{P}^{(r)}$ and using the average difference $\Delta_{\boldsymbol{\theta}}^{(r)} = \frac{1}{P}\sum_{j\in \mathcal{P}^{(r)}}(\hat{\boldsymbol{\theta}}_j^{(r)} - \boldsymbol{\theta}_j^{(r)})$ as the shared update.
    \item \textit{Task Arithmetic}: Applying this update to all model copies independently $\tilde{\boldsymbol{\theta}}_j^{(r+1)} = \boldsymbol{\theta}_j^{(r)} +\Delta_{\boldsymbol{\theta}}^{(r)}$, $\forall j\in \mathcal{N}$.
    \item \textit{Collusion-aware Embedding}: Training all model copies $\tilde{\boldsymbol{\theta}}_j^{(r+1)}$ on the current version of the shared trigger set $\mathcal{X}^{(r)}$ with their respective labels $\mathbf{y}_j$ to achieve the watermarked model copies $\boldsymbol{\theta}_j^{(r+1)}$. 
    \item \textit{Optimization of the Shared Trigger Set}: Updating the trigger set $\mathcal{X}^{(r)}$ according to the current model copies to achieve $\mathcal{X}^{(r+1)}$.\footnote{The impact of a potential mismatch between the leak model round and trigger set version is evaluated in Section~\ref{sec:exp}.}
    \item \textit{Return}: Transferring model copies $\boldsymbol{\theta}_j^{(r+1)}$ back to each data-owner for the next round.
\end{enumerate}

\subsection{Collusion-Aware Embedding}

Independently embedding a unique watermark $\mathcal{W}_j = \{\mathbf{X},\mathbf{y}_j\}$ into the different model copies $\boldsymbol{\theta}_j$ does not, by itself, guarantee robustness to a collusion attack. We show in Figure~\ref{fig:collusion_comparison} several methods to degrade the trigger accuracy, where the simple average of the parameters of two models (i.e., a collusion attack by two data-owners) is the most disruptive modification and also manages to maintain a high task accuracy. To address this, we propose a collusion-aware term added into the embedding loss function, taking advantage of the aggregator's white-box access to all model copies. At the embedding step for each data-owner $j$, the aggregator can sample $\mathcal{N}$ to select a subset of $M$ auxiliary data-owners $\mathcal{M}^{(j)}\subseteq\mathcal{N}\backslash\{j\}$, and emulate virtual collusions to optimize their resulting MAV,\footnote{We consider balanced collusions for simplicity and robustness, as a weighted average would always expose one of the colluders more on the resulting merged model. Section~\ref{sec:exp} also analyzes the impact of attacking with an alternative collusion implementation.} minimizing 
\begin{equation} \label{eq:colloss}
L^\text{CA}_j =  \sum_{m\in \mathcal{M}^{(j)}} \text{min}\left( \text{CE}(\mathcal{W}_j,\boldsymbol{\theta}_{\{j,m\}} ),\text{CE}(\mathcal{W}_m,\boldsymbol{\theta}_{\{j,m\}} )\right),
\end{equation}
where
\begin{equation}\label{eq:averaging}
    \boldsymbol{\theta}_{\{j,m\}} = \frac{1}{2} (\boldsymbol{\theta}_j + \boldsymbol{\theta}_m),
\end{equation}
$\text{CE}(\mathcal{W},\boldsymbol{\theta})$ represents the cross entropy loss of dataset $\mathcal{W}$ on the model with parameters $\boldsymbol{\theta}$, and the $\text{min}(\cdot)$ function guarantees the chosen collusion label does not contradict previous rounds, or the inherent trends of the resulting model. The overall embedding loss for the $j$th model is
\begin{equation}
    L_j =  \text{CE}(\mathcal{W}_j,\boldsymbol{\theta}_j) + \lambda_\text{CA} \cdot L^\text{CA}_j,
\end{equation}
with a parameter $\lambda_\text{CA}$ controlling the strength of the collusion-aware loss. 

\subsection{Optimization of the Shared Trigger Set}

The unique embedding into the different model copies enables collusion-resistant traitor tracing, but its effects on the FL dynamics introduce a challenging trade-off between a fast memorization and the undesirable impact of large parameter heterogeneity within the system, which can result in unhelpful or even destructive shared updates. To alleviate these constraints, the aggregator can alternately update the model parameters and optimize the shared trigger set $\mathcal{X}$ through $K$ adversarial perturbation steps, enhancing the features such that more subtle differences between model copies are enough to classify the triggers with their assigned labels. While various adversarial example generation mechanisms could be used (e.g., CW~\cite{CW}, PGD~\cite{PGD}, FGSM~\cite{FGSM}, etc.), we adopt the Projected Gradient Descent (PGD) into the valid image space, for simplicity:
\begin{equation}
    \mathbf{x}^{(r,k+1)} = \mathbf{x}^{(r,k)} + \epsilon \cdot \text{sign}\left(\nabla_{\mathbf{x}^{(r,k)}} \sum_{j\in \mathcal{N}} L_j\right),
\end{equation}
ensuring that
\begin{equation}
    ||\mathbf{x}^{(r,k+1)} - \mathbf{x}^{(0)}||_{\infty}\leq \alpha,
\end{equation}
for $k=0,...,K-1$, where $\epsilon$ is the adversarial step size, and $\alpha$ is the perturbation budget. After $K$ optimization steps, the best trigger set is chosen that minimizes the embedding loss: 
\begin{equation}
\mathbf{x}^{(r+1)} = 
\underset{k = 0,\ldots,K}{\arg\min} \;
\sum_{j \in \mathcal{N}} L_j\big(\mathbf{x}^{(r,k)}\big).
\end{equation}

\subsection{Functional Regularization against Model Drift}

Traitor tracing inherently requires numerically different model copies for each data-owner, yet such parameter differences can negatively impact the compatibility of shared model updates, leading to poor convergence and unstable training. This effect becomes more evident in architectures with Batch Normalization (BN) layers. Similar to the case of data heterogeneity~\cite{Guerraoui24}, the heterogeneity in model parameters produces different feature distributions that cannot be aligned with the same normalization. Over time, because each iteration does not unify all model copies, these differences cause them to drift apart in the parameter space, resulting in ineffective or even destructive shared updates across model copies. To mitigate this effect, we propose BlackCATT+FR, where an additional \textit{functional regularization} term $L^\text{FR}$ is introduced to preserve the shared feature space across model copies through a shared reference sampled from an auxiliary dataset $\mathcal{D}_\text{aux}$ at the aggregator. These auxiliary samples serve as anchors for natural images, in contrast to the artificial nature of the shared trigger set $\mathcal{X}$ where model copies are expected to exhibit different behavior. The auxiliary dataset does not need to be labeled, class-balanced or even task-related (see Section~\ref{sec:exp:abl}), which makes this approach still feasible for an aggregator without explicit knowledge of the main task. Considering this, the total loss in BlackCATT+FR becomes
\begin{multline}
    L_j =  \text{CE}(\mathcal{W}_j,\boldsymbol{\theta}_j) + \lambda_\text{CA} \cdot L^\text{CA}_j + \lambda_\text{FR} \cdot L^\text{FR}_j, \\ \text{with} \quad
   L^\text{FR}_j = \mathbb{E}_{\boldsymbol{x}\sim\mathcal{D}_\text{aux}}\left[D_{\text{KL}}(f_{\boldsymbol{\theta}^{(r)}_j}(\boldsymbol{x}) \| f_{\bar{\boldsymbol{\theta}}_j^{(r)}}(\boldsymbol{x}))\right],
\end{multline}
where
\begin{equation}
   \bar{\boldsymbol{\theta}}^{(r)} = \frac{1}{N} \sum_{j\in\mathcal{N}} \boldsymbol{\theta}_j^{(r)}
\end{equation}
denotes the global average of all model copies approximating the shared main task function, $D_{\text{KL}}$ is the Kullback-Leibler divergence to minimize the distance between the soft outputs in both functions, and $\lambda_\text{FR}$ is a parameter controlling the strength of the regularization.

\section{Experimental Analysis} \label{sec:exp}

\subsection{Experimental Settings}
\subsubsection{Models and Datasets}
The proposed scheme has been tested on two popular architectures (VGG16~\cite{VGG} and ResNet18~\cite{ResNet}) and datasets (CIFAR-10 and CIFAR-100~\cite{CIFAR10}). Unless stated otherwise, the default configuration for the experiments is ResNet18 with CIFAR-100. 

\subsubsection{Federated Learning Settings}
The default number of data-owners considered is $N = 20$, with only $P = 10$ called to participate in each round. All experiments start from a random initialization, with the local main task training using a batch size of 64 for a full data-owner epoch, a learning rate $lr_\text{MT} = 0.01$, and Stochastic Gradient Descent with momentum $0.9$ and weight decay of $10^{-4}$. To mitigate overfitting, data-owners employ data augmentation (random crops with 4-pixel padding and horizontal flips).

\subsubsection{Watermarking Configuration}
Throughout these experiments, the watermark embedding uses a much lower learning rate than for the main task $lr_\text{WM} =10^{-4}$ with the same kind of optimizer, and trains once per round on a single batch containing the full trigger set $\mathcal{X}$ (default $T = 250$). BN layers are frozen during embedding~\cite{Shao23}.\footnote{To avoid degradation of the main task.}  Unless stated otherwise, the trigger set $\mathcal{X}$ is initialized from a discrete uniform distribution of valid pixel values. When used, the collusion-aware loss adopts $\lambda_\text{CA} = 0.1$ and $M = 5$. Adversarial trigger optimization allows a default perturbation budget $\alpha = 64$, step size $\epsilon = 1$, and $K=1$. Functional regularization uses an auxiliary dataset $\mathcal{D}_\text{aux}$, with no relation to the main task, of 500 WikiArt~\cite{WikiArt} samples in batches of size 64, and $\lambda_\text{FR} = 0.1$. For Tardos Codes, we set $\epsilon_{\text{FP}} =10^{-6}$, $\kappa = 0.5$\footnote{A higher value of $\kappa =100$ was found to be more beneficial for DNN traitor tracing in \cite{RodriguezLois23}, as it is more suitable for a collusion resembling a majority voting scenario. However, values greater than $\kappa = 0.5$ can increase the probability of false positives in Tardos Codes \cite{Simone11}. Although this probability is theoretically bounded by~\eqref{eq:quad} under the assumption of independent triggers,  trigger set optimization may weaken this assumption. Experimentally, using $\kappa=100$ resulted in non-negligible false accusations against specific users in certain cases, particularly for VGG16 when a colluded model was sufficiently pruned and later re-trained, allowing it to recover main-task biases while mostly removing watermark information.} and the cutoff parameter is set according to the number of classes $q$, ($\tau_{q=10}= 0.01$, $\tau_{q=100}= 0.001$). 

Table~\ref{tab:schemes} summarizes the evaluated configurations. While there is no established state of the art for collusion-resistant black box traitor tracing, Vanilla embedding is comparable to the setups used in previous works such as~\cite{Fang-Qi_Li21,RodriguezLois24}. Although they present small variations, these are not particularly relevant for the current comparison. The method in~\cite{RodriguezLois24} adds dropout layers to the shallow architecture considered; however, experimental tests showed this did not allow collusion-resistant embedding in larger networks. In~\cite{Fang-Qi_Li21}, each data owner receives a unique trigger–label pair, which were shown to still be vulnerable to collusion in the same way as a shared trigger set~\cite{RodriguezLois24}. The triggers in~\cite{Fang-Qi_Li21} are also designed to match the distribution of the data-owner's examples, which offers advantages in terms of stealthiness but is expected to increase the MAV, in favor of the task's correct label~\cite{RodriguezLois23}.

\begin{table}[ht]

\caption{Configuration of the different embedding approaches compared in the experimental section.}\label{tab:schemes}
\centering
\resizebox{\linewidth}{!}{
\begin{tabular}{r|ccc||cc|}
\cline{2-6}
& \multicolumn{3}{c||}{Loss Term}                                        & \multicolumn{2}{c|}{Optimizing}                              \\ \cline{2-6} 
                                                              & \multicolumn{1}{c|}{CE } & \multicolumn{1}{c|}{CA } & FR  & \multicolumn{1}{c|}{$\boldsymbol{\theta}_j$} & $\mathcal{X}$ \\ \hline
\multicolumn{1}{|r||}{No WM}                                   & \multicolumn{1}{c|}{}        & \multicolumn{1}{c|}{}        &         & \multicolumn{1}{c|}{}                        &               \\ \hline
\multicolumn{1}{|r||}{Vanilla}                                 & \multicolumn{1}{c|}{\checkmark}       & \multicolumn{1}{c|}{}        &         & \multicolumn{1}{c|}{\checkmark}                       &               \\ \hline
\multicolumn{1}{|r||}{BlackCATT}                                    & \multicolumn{1}{c|}{\checkmark}       & \multicolumn{1}{c|}{\checkmark}       &         & \multicolumn{1}{c|}{\checkmark}                       & \checkmark             \\ \hline
\multicolumn{1}{|r||}{BlackCATT+FR}                                 & \multicolumn{1}{c|}{\checkmark}       & \multicolumn{1}{c|}{\checkmark}       & \checkmark       & \multicolumn{1}{c|}{\checkmark}                       & \checkmark             \\ \hline
\multicolumn{1}{|r||}{BlackCATT+FR w/o $\nabla_{\mathcal{X}^{(r)}}$} & \multicolumn{1}{c|}{\checkmark}       & \multicolumn{1}{c|}{\checkmark}       & \checkmark       & \multicolumn{1}{c|}{\checkmark}                       &               \\ \hline
\multicolumn{1}{|r||}{BlackCATT+FR w/o $L^\text{CA}$}               & \multicolumn{1}{c|}{\checkmark}       & \multicolumn{1}{c|}{}        & \checkmark       & \multicolumn{1}{c|}{\checkmark}                       & \checkmark             \\ \hline
\end{tabular}
}
\end{table}

\subsubsection{Attack Configuration}
The following attacks are considered in this section:
\begin{itemize}
    \item Collusion attacks:
    \begin{itemize}
        \item Parameter averaging: Average of model parameters, as defined in~\eqref{eq:averaging}.
        \item Random layer sampling: Randomly selecting full layers from different data-owners.
    \end{itemize}
    \item Further attacks on the colluded model (fine-pruning):
    \begin{itemize}
        \item Pruning: Zero-ing out the desired ratio of parameters with the lowest L1-norm.
        \item Fine-tuning: Performing 5 epochs with training data from all malicious participants and $lr_\text{FT} = 0.01$.
    \end{itemize}
    
\end{itemize}

\subsubsection{Experimental metrics}
We evaluate each scheme over 100 random collusions, except for fine-tuning experiments, which were only repeated over 50 random collusions due to computational constraints. Traitor tracing performance is measured by the resulting MAV~\eqref{eq:mav} for a collusion $\mathcal{C}^2$, the average number of exposed triggers $t^*$ before an accusation, and the False Negative Rate (FNR) after querying the full trigger set $\mathcal{X}$. Main task impact and convergence are assessed through the overall accuracy on a centralized test set of 10,000 images (unless stated otherwise). 

\subsection{FL Convergence}\label{sec:exp:fl}

\begin{figure}[ht]
    \centering
    \subfloat{\includegraphics[height=0.11\columnwidth]{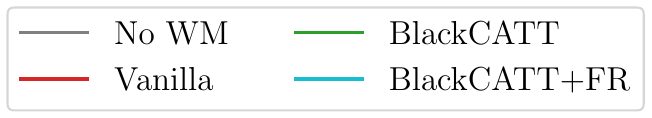}}\\
    \subfloat[VGG16 on CIFAR-10]{\includegraphics[width=\columnwidth]{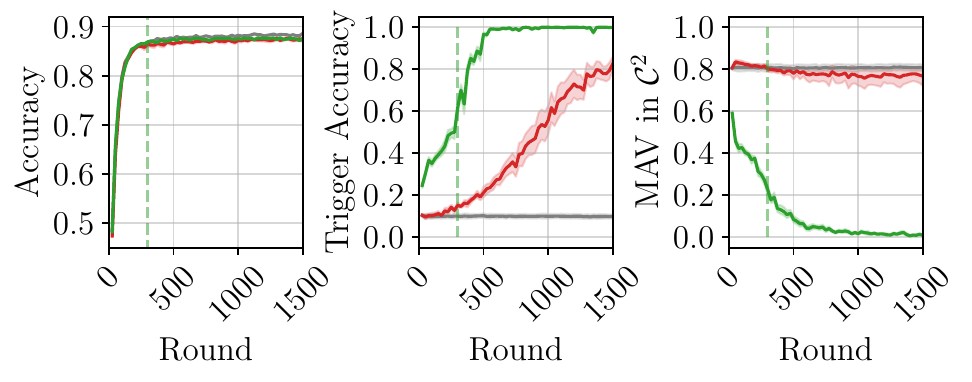}}\\
    \subfloat[ResNet18 on CIFAR-100]{\label{fig:evol:b}\includegraphics[width=\columnwidth]{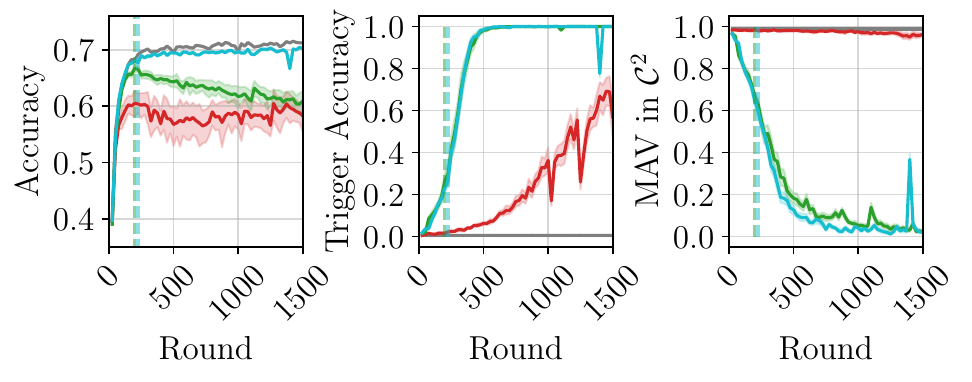}}
    \caption{Evolution of training metrics, with a vertical dashed line representing the FNR $\simeq 0.5$ threshold for $\mathcal{C}^2$.}
    \label{fig:evol}
\end{figure}

Figure \ref{fig:evol} shows the training evolution across FL rounds, comparing the convergence of the main and watermarking tasks. As mentioned in Section \ref{sec:catt}, the detrimental effect of the unique watermark embedding is much more noticeable in ResNet18 (VGG16 does not have BN layers), but this can be mitigated with the addition of functional regularization (BlackCATT+FR). The proposed BlackCATT serves a dual purpose; it allows for more subtle parameter differences between model copies than Vanilla embedding, which improves main task dynamics; and significantly accelerates watermark convergence, protecting model copies from the early stages of training. Traitor tracing works often consider an FNR $\simeq 0.5$ to be a sufficient deterrent for potential colluders~\cite{Skoric12}, so this threshold is highlighted for a collusion $\mathcal{C}^2$ in the Figure, to indicate the point from which the model is considered protected. Of course, potential colluders cannot evaluate the probability of detection at any given point, and earlier leaks would still imply a non-negligible probability of being caught.

\subsubsection{Impact of the number of data-owners}
\begin{figure}[ht]
    \centering
    \subfloat{\includegraphics[height=0.07\columnwidth]{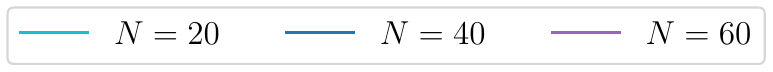}}\\
    \subfloat{\includegraphics[width=\columnwidth]{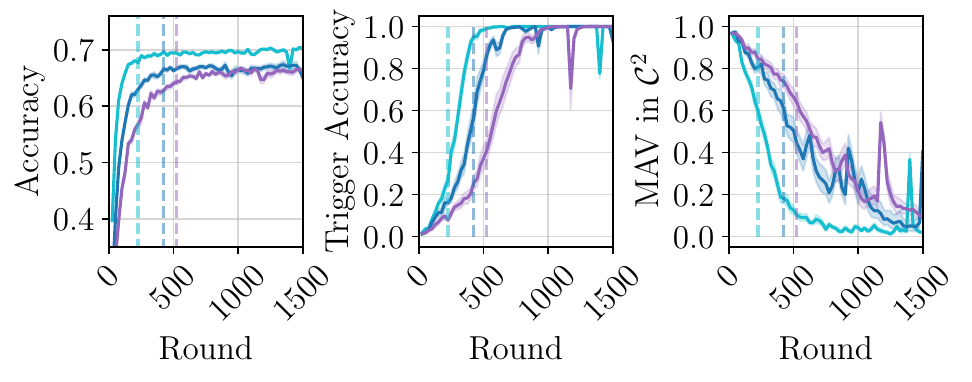}}
    
    \caption{Impact of the number of data-owners $N$, with a vertical dashed line representing the FNR $\simeq 0.5$ threshold for $\mathcal{C}^2$.}
    \label{fig:impact_n}
\end{figure}

A higher number of data-owners slows the convergence of both the main task and the watermark embedding but, as can be seen in Figure~\ref{fig:impact_n}, this does not compromise the effectiveness of the scheme. Although the detection threshold is delayed, it also aligns with a lower-accuracy round for larger number of data-owners, albeit at the expense of a higher computational cost on the aggregator, as the number of model copies grows.

\subsection{Analysis of the Attack}

\begin{figure}[ht]
    \centering
    \subfloat{\includegraphics[height=0.065\columnwidth]{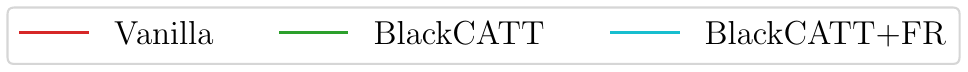}}\\
    \subfloat[VGG16 on CIFAR-10]{\label{fig:tt:a}\includegraphics[width=\columnwidth]{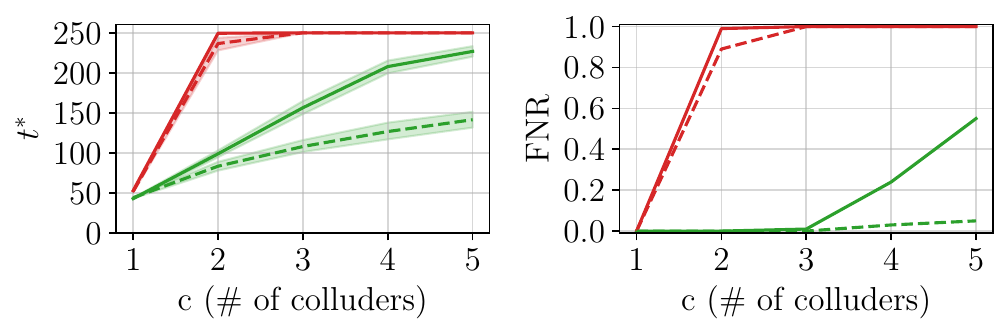}}\\
    \subfloat[ResNet18 on CIFAR-100]{\label{fig:tt:b}\includegraphics[width=\columnwidth]{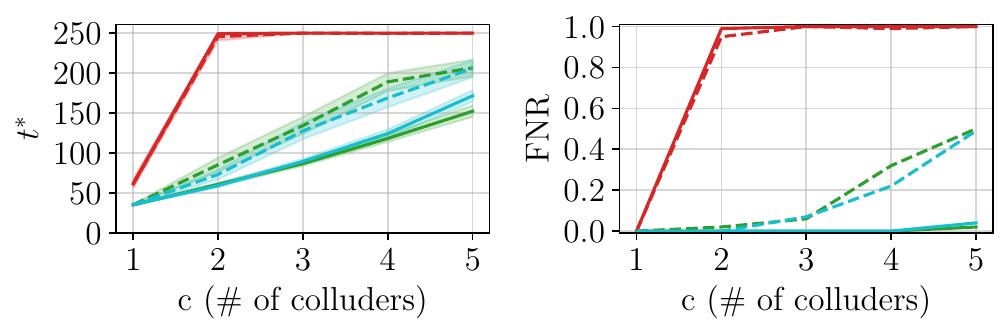}}\\
    \caption{Traitor tracing capabilities against parameter averaging (solid line) and random layer sampling (dashed line) in terms of exposed triggers before an accusation ($t^*$) and FNR.}
    \label{fig:tt}
\end{figure}

 As indicated by the MAV remaining high in Figure~\ref{fig:evol}, Vanilla embedding into the model copies is not sufficient to achieve collusion-resistant traitor tracing, but this becomes even more evident in Figure~\ref{fig:tt}, comparing the traitor tracing performance of the different approaches after a potential model leak at the end of training.  Even though the identification of a single malicious participant is comparable to that of the proposed approach, the Vanilla embedding fails to find the culprits if there is a collusion among data-owners. Furthermore, while the collusion-aware loss $L^\text{CA}$ samples random collusions through parameter averaging, the scheme remains mostly robust to a different collusion implementation (random layer sampling), as shown in the dashed lines in Figure~\ref{fig:tt}.

\subsubsection{Impact of further attacks}

\begin{figure}[ht]
    \centering
    \subfloat{\includegraphics[height=0.07\columnwidth]{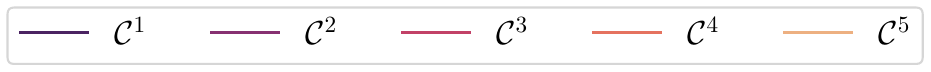}}\\
    \subfloat[VGG16 on CIFAR-10]{\label{fig:ttfp:a}\includegraphics[width=\columnwidth]{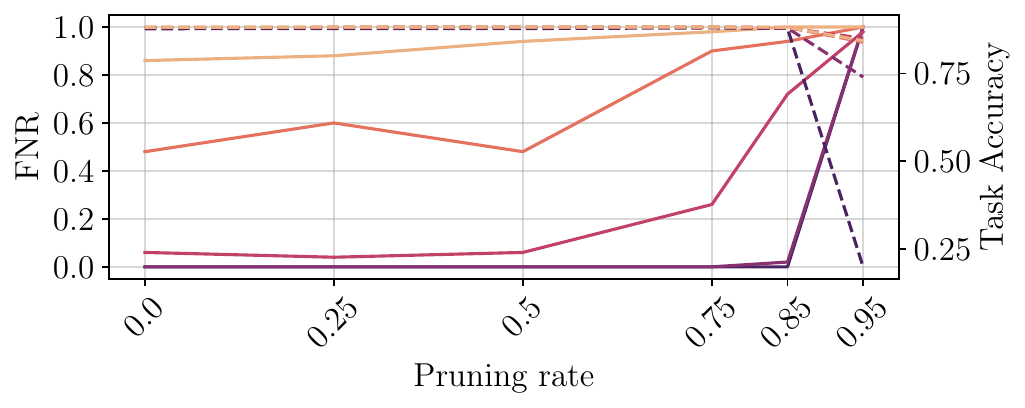}} \\
    \subfloat[ResNet18 on CIFAR-100]{\label{fig:ttfp:b}\includegraphics[width=\columnwidth]{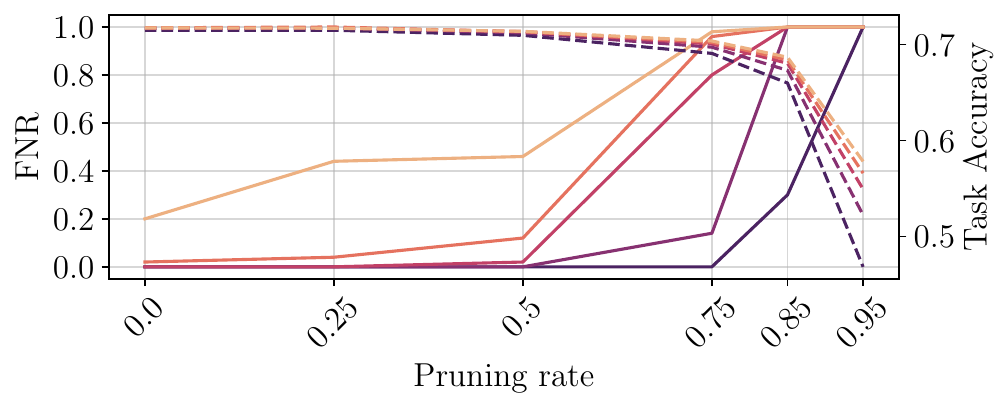}}\\
    \caption{Traitor tracing FNR (solid line) and main task accuracy (dashed line) against model averaging for different collusion sizes with pruning and fine-tuning of the merged model.}
    \label{fig:ttfp}
\end{figure}

While this work focuses on the collaborative nature of collusion attacks, it is reasonable to expect a malicious participant would attempt to remove the watermark from the colluded model through additional efforts, especially if they are able to utilize their respective training datasets. Considering fine-pruning with all colluders' data, after pruning the merged parameters with different pruning ratios, Figure~\ref{fig:ttfp} showcases the effect of these attacks: as the FNR increases with stronger pruning, the accuracy on the main task decreases, suggesting that the proposed watermark is not easily removable from the network. On the other side, a higher number of colluders with better fine-tuning capabilities becomes more challenging to detect, but detection remains likely up to $\mathcal{C}^3$ for CIFAR-10 and $\mathcal{C}^4$ for CIFAR-100. Importantly, even under strong attack settings, colluders have no reliable way to estimate their likelihood of detection, which maintains the deterrent effect of the scheme.

\subsubsection{Leaking an earlier training round}
Considering the convergence of the main task and traitor tracing in Figure~\ref{fig:evol}, a malicious participant may choose to steal a suboptimal but less protected version of their model copy. As the optimization of the shared trigger set $\mathcal{X}^{(r)}$ progresses, a mismatch between the version of the trigger set and the unknown leak round could negatively affect detection performance. Table~\ref{tab:mismatch} contains the resulting FNR for mismatched detection attempts of 2 ($\mathcal{C}^2$) and 5 ($\mathcal{C}^5$) malicious participants. While the latest version of the trigger set $\mathcal{X}^{(1500)}$ is effective for most potential leaks, this mismatch could present problems in applications with fast main task convergence. To prevent this, the trigger optimization could be frozen in earlier rounds, or strengthened through additional computation (see Section~\ref{sec:exp:t_optim}). 

\begin{table}[ht]
\centering
\caption{Traitor tracing effectiveness (FNR) against the leak of an earlier, less accurate version of the model, querying different versions of the trigger set $\mathcal{X}^{(r)}$, for $\mathcal{C}^2$ ($\mathcal{C}^5$).}\label{tab:mismatch}
\begin{tabular}{r|cccc|}
\cline{2-5}
\multicolumn{1}{l|}{}& \multicolumn{4}{c|}{Trigger Set Version $(r)$}\\ 
\cline{2-5}
\multicolumn{1}{l| }{} &  \multicolumn{1}{c|}{250} & \multicolumn{1}{c|}{500} & \multicolumn{1}{c|}{1000} & 1500\\ 
\hline \multicolumn{1}{|c|}{Round (Acc.)}&\multicolumn{4}{c|}{VGG16 on CIFAR-10} \\ \hline
 \multicolumn{1}{|r|}{250 (0.865)}& \multicolumn{1}{c|}{0.63 (1.00)}& \multicolumn{1}{c|}{0.95 (1.00)}& \multicolumn{1}{c|}{1.00 (1.00)}&1.00 (1.00)\\ \hline
 \multicolumn{1}{|r|}{500 (0.875)}&  \multicolumn{1}{c|}{0.82 (0.99)}& \multicolumn{1}{c|}{\textbf{0.00} (0.50)}& \multicolumn{1}{c|}{0.10 (0.66)}&0.20 (0.94)\\ \hline
 \multicolumn{1}{|r|}{1000 (0.876)}& \multicolumn{1}{c|}{0.93 (1.00)}& \multicolumn{1}{c|}{\textbf{0.00} (0.91)}& \multicolumn{1}{c|}{\textbf{0.00} (0.55)}&\textbf{0.00} (0.81)\\ \hline
 \multicolumn{1}{|r|}{1500 (\textbf{0.876})}& \multicolumn{1}{c|}{0.95 (1.00)}& \multicolumn{1}{c|}{0.01 (0.96)}& \multicolumn{1}{c|}{\textbf{0.00} (0.50)}&\textbf{0.00} (0.54)\\ \hline
\multicolumn{1}{l|}{} & \multicolumn{4}{c|}{ResNet18 on CIFAR-100}\\ \hline
 \multicolumn{1}{|r|}{250 (0.689)}& \multicolumn{1}{c|}{0.08 (0.82)}& \multicolumn{1}{c|}{0.78 (1.00)}& \multicolumn{1}{c|}{0.87 (1.00)}&0.72 (1.00)\\ \hline
 \multicolumn{1}{|r|}{500 (0.694)}&  \multicolumn{1}{c|}{0.02 (0.94)}& \multicolumn{1}{c|}{\textbf{0.00} (0.13)}& \multicolumn{1}{c|}{\textbf{0.00} (0.36)}&\textbf{0.00} (0.39)\\ \hline
 \multicolumn{1}{|r|}{1000 (0.697)}& \multicolumn{1}{c|}{0.02 (0.98)}& \multicolumn{1}{c|}{\textbf{0.00} (0.51)}& \multicolumn{1}{c|}{\textbf{0.00} (0.01)}&\textbf{0.00} (0.11)\\ \hline
 \multicolumn{1}{|r|}{1500 (\textbf{0.702})}& \multicolumn{1}{c|}{\textbf{0.00} (0.96)}& \multicolumn{1}{c|}{\textbf{0.00} (0.58)}& \multicolumn{1}{c|}{\textbf{0.00} (0.12)}&\textbf{0.00 (0.00)}\\ \hline

\end{tabular}
\end{table}

\subsection{Parameters for the Defense}

\subsubsection{Number of triggers}

\begin{figure}[ht]
    \centering
    \subfloat{\includegraphics[height=0.07\columnwidth]{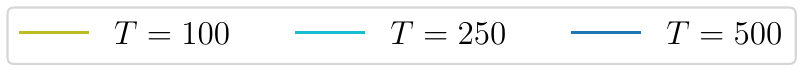}}\\
    \subfloat{\includegraphics[width=\columnwidth]{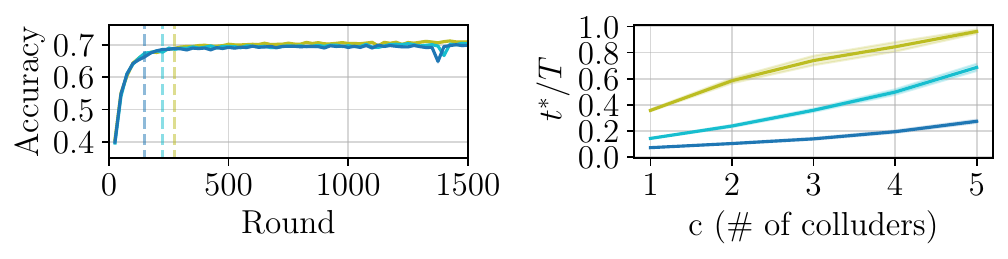}}
    
    \caption{Impact of the number of triggers on task convergence, with a vertical dashed line representing the FNR $\simeq 0.5$ threshold for $\mathcal{C}^2$, and traitor tracing performance as the ratio of exposed triggers $t^* /T$ before an accusation.}
    \label{fig:impact_m}
\end{figure}

Figure \ref{fig:impact_m} presents the training evolution and the average ratio $t^*/T$ for different sizes $T$ of the trigger set. While more computationally expensive, larger trigger sets offer more opportunities to query a model, with two main benefits: they allow detection in earlier stages when the MAV is higher and some queries are not yet useful, and, in later stages, they preserve unused triggers from spoil attacks~\cite{Fang-Qi_Li21} to be used in future accusations.

\subsubsection{Trigger optimization rounds} \label{sec:exp:t_optim}
\begin{figure}[ht]
    \centering
    \includegraphics[width=\columnwidth]{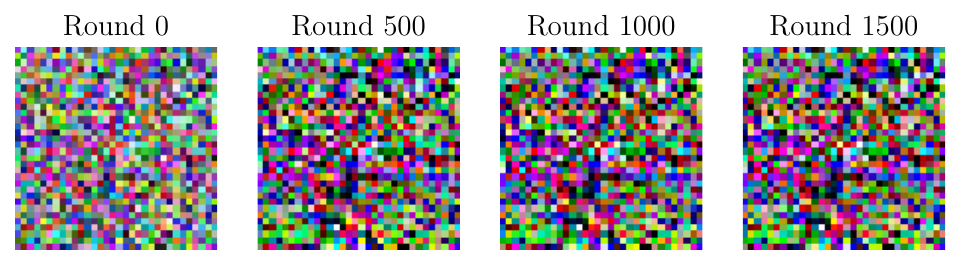}
    \caption{Evolution of a specific trigger $\mathbf{x} \in \mathcal{X}$ during training.}
    \label{fig:trigger}
\end{figure}

The adversarial optimization of the shared trigger set $\mathcal{X}$ can be computationally expensive, as it jointly optimizes the embedding loss across all model copies to converge toward features that enhance traitor tracing, as shown in the example in Figure~\ref{fig:trigger}. However, in cases where the aggregator is able to perform multiple optimization iterations $K$ per training round, the performance of the scheme improves significantly, as noted in Table~\ref{tab:impact_rounds}. With this faster traitor tracing convergence, model copies are protected from earlier rounds with lower main task accuracy, and the impact of potential early leaks is mitigated, without substantial degradation of main task performance.

\begin{table}[ht]
\centering
\caption{Impact of the number of trigger set $\mathcal{X}$ optimization iterations $K$ per training round, on the traitor tracing convergence and FNR under round leak mismatch.}\label{tab:impact_rounds}
\begin{tabular}{r|ccc|}
\cline{2-4}
\multicolumn{1}{l|}{}& \multicolumn{3}{c|}{Trigger Optim. Iterations ($K$) }\\ 
\cline{2-4}
\multicolumn{1}{l| }{} & \multicolumn{1}{c|}{1} & \multicolumn{1}{c|}{2} & 5\\ \hline
 \multicolumn{1}{|r|}{Round @ FNR $\simeq 0.5$ for $\mathcal{C}^2$}& \multicolumn{1}{c|}{255}& \multicolumn{1}{c|}{175}& \textbf{125}\\ \hline
 \multicolumn{1}{|r|}{Acc. @ FNR $\simeq 0.5$ for $\mathcal{C}^2$}& \multicolumn{1}{c|}{0.677}& \multicolumn{1}{c|}{0.670}& \textbf{0.649}\\ \hline
 \multicolumn{1}{|r|}{Max. Accuracy}& \multicolumn{1}{c|}{\textbf{0.705}}& \multicolumn{1}{c|}{ 0.703}& 0.695\\ \hline
 \multicolumn{1}{|r|}{$\mathcal{C}^2$ FNR $\boldsymbol{\theta}_\mathcal{C}^{(250)}$, $\mathcal{X}^{(1500)}$}& \multicolumn{1}{c|}{ 0.72 }& \multicolumn{1}{c|}{ 0.69}& \textbf{0.43} \\ \hline

\end{tabular}
\end{table}

\subsubsection{Trigger stealthiness}
\begin{figure}[ht]
    \centering
    \includegraphics[width=\columnwidth]{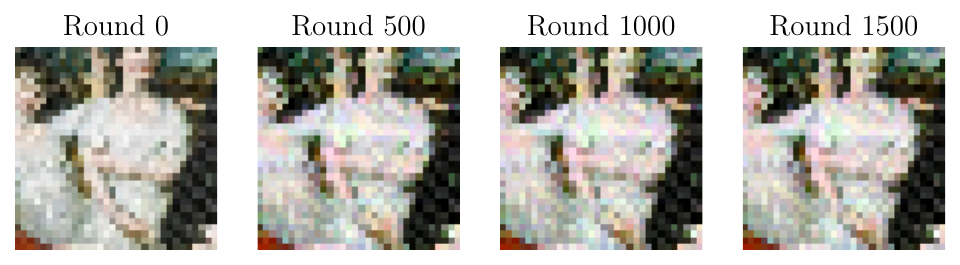}
    \caption{Evolution of a stealthy trigger $\mathbf{x} \in \mathcal{X}$ during training.}
    \label{fig:trigger_stealth}
\end{figure}

\begin{figure}[ht]
    \centering
    \subfloat{\includegraphics[height=0.065\columnwidth]{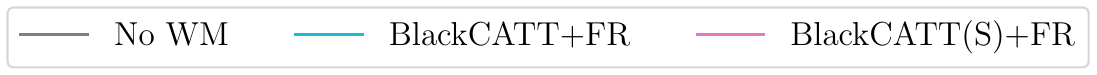}}\\
    \subfloat{\includegraphics[width=\columnwidth]{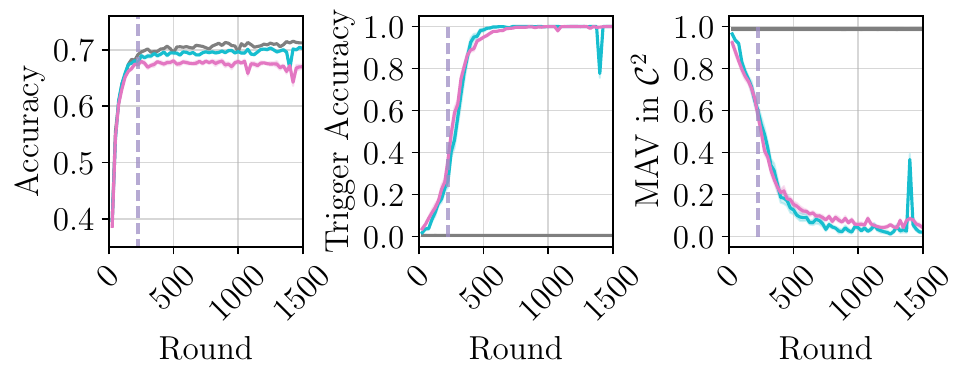}}
    \caption{Comparison of the training metrics for stealthy triggers across training rounds, with a vertical dashed line representing the FNR $\simeq 0.5$ threshold for $\mathcal{C}^2$.}
    \label{fig:stealth}
\end{figure}

The proposed adversarial optimization of the shared trigger set $\mathcal{X}$ promotes collusion-resistant features starting from an initial set $\mathcal{X}^{(0)}$, which across this work has been initialized as a random sample from a discrete uniform distribution, as shown in Figure~\ref{fig:trigger}. While this initialization works well for traitor tracing, suspicious queries depicting only noise could potentially be detected and bypassed by a knowledgeable adversary. To assess the potential impact of using less obvious triggers, we test BlackCATT(S)+FR, initializing the shared trigger set $\mathcal{X}^{(0)}$ as natural images from the same source as $\mathcal{D}_\text{aux}$, and lowering the perturbation budget $\alpha=16$. An example of the evolution of a single trigger is shown in Figure~\ref{fig:trigger_stealth}, which would be stealthier than the random noise in Figure~\ref{fig:trigger}. However, considering the results in Figure~\ref{fig:stealth}, the choice of stealthy triggers does come at a slight cost of model performance, potentially due to the closer distribution to the main task datasets, and presents higher values of MAV, making them less efficient for traitor tracing.  

\subsection{Ablation Studies} \label{sec:exp:abl}

\subsubsection{Component ablation analysis}
\begin{figure}[ht]
    \centering
    \subfloat{\includegraphics[height=0.11\columnwidth]{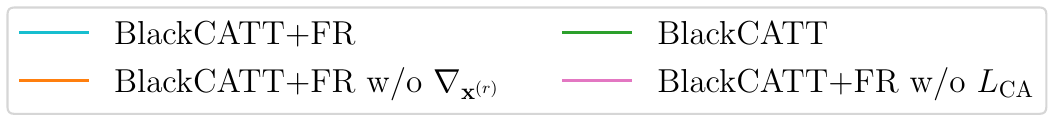}}\\
    \subfloat{\includegraphics[width=\columnwidth]{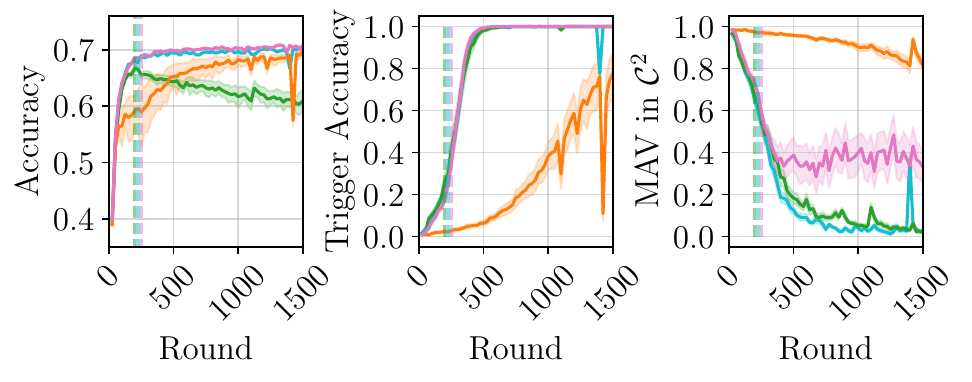}}
    
    \caption{Impact of eliminating the different components of the proposed scheme, with a vertical dashed line representing the FNR $\simeq 0.5$ threshold for $\mathcal{C}^2$.}
    \label{fig:ablation}
\end{figure}

Figure~\ref{fig:ablation} compares the proposed BlackCATT+FR with the ablated versions defined in Table~\ref{tab:schemes}. The effect of each of these components is evident: BlackCATT+FR w/o $\nabla_{\mathcal{X}^{(r)}}$ presents a much slower convergence of both tasks, especially the watermark, as it requires more evident differences between model copies to correctly classify the trigger set; BlackCATT+FR w/o $L^\text{CA}$ still provides significant convergence speed, but in turn is not too efficient for traitor tracing, as the MAV is not explicitly optimized; and finally BlackCATT which cannot compensate for model drift, as discussed in Section~\ref{sec:exp:fl}.

\subsubsection{Impact of auxiliary aggregator dataset}
\begin{figure}[ht]
    \centering
    \subfloat[CIFAR-100]{\includegraphics[width=\columnwidth]{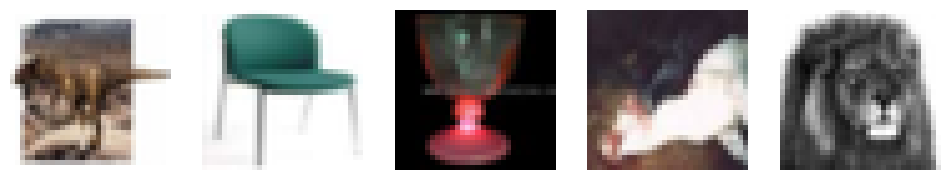}}\\
    \subfloat[Tiny ImageNet]{\includegraphics[width=\columnwidth]{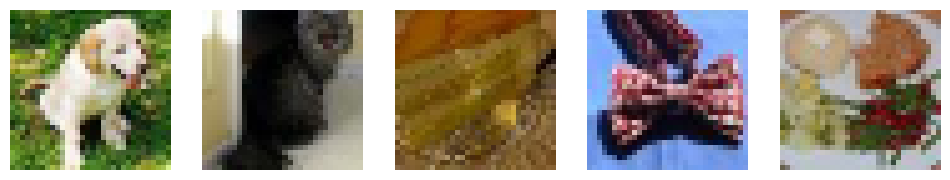}}\\
    \subfloat[WikiArt]{\includegraphics[width=\columnwidth]{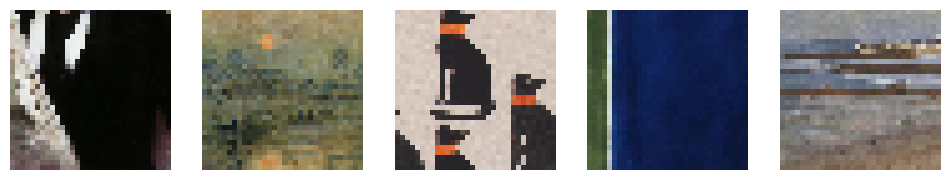}}\\
    
    \caption{Examples of images from the different $\mathcal{D}_\text{aux}$ datasets.}
    \label{fig:aux_datasets}
\end{figure}

\begin{figure}[ht]
    \centering
    \subfloat{\includegraphics[height=0.06\columnwidth]{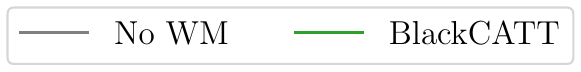}}\\
    \subfloat{\includegraphics[height=0.06\columnwidth]{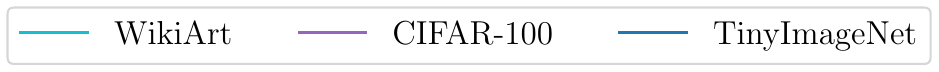}}\\
    \subfloat{\includegraphics[width=\columnwidth]{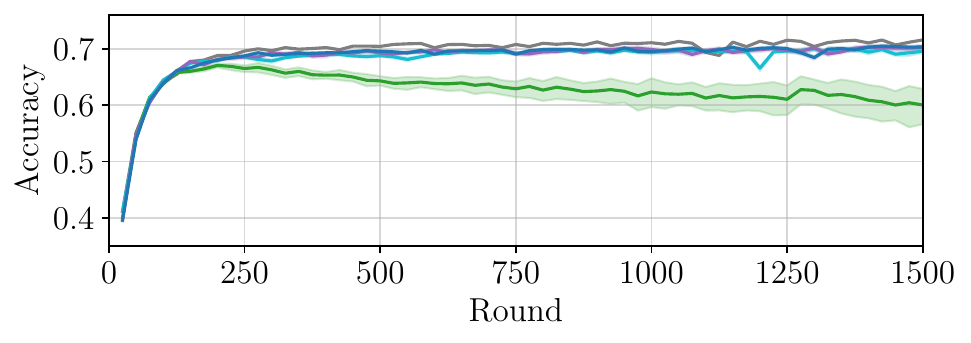}}
    
    \caption{Impact of different $\mathcal{D}_\text{aux}$ datasets in BlackCATT+FR.}
    \label{fig:impact_aux_dataset}
\end{figure}

Providing the aggregator with an auxiliary, unlabeled dataset to perform functional regularization could be impractical in a real-world application. Because of this, the impact of dataset choice $\mathcal{D}_\text{aux}$ was studied: matching the main task (CIFAR-100),\footnote{The auxiliary samples at the aggregator were taken from the test set, and were not used in training or testing for any approach in this comparison.} using a dataset with a similar distribution (Tiny ImageNet~\cite{TinyImageNet}), and one with a substantial task and resolution difference (WikiArt~\cite{WikiArt}). Samples from these datasets can be seen in Figure \ref{fig:aux_datasets}.\footnote{Already adjusted to the input dimensions ($32\times 32$) through cropping and resizing.} When compared to the No WM reference, and the proposed BlackCATT in Figure~\ref{fig:impact_aux_dataset}, it appears the choice of dataset is not critical, meaning the aggregator would not need highly accurate information about the main task to perform functional regularization.

\subsection{Experimental analysis of the False Positive Rate}\label{sec:exp:expfpr}

The theoretical analysis of Tardos codes in Equation~\eqref{eq:quad} assumes triggers are independent of their assigned labels. In practice, however,  this assumption may no longer hold, due to the complexity of the learned function and the adversarial optimization used, as the trigger set $\mathcal{X}$ could potentially learn label-specific features that improve classification loss. Although even then, a false accusation would require a data-owner to be assigned labels matching these classifications trends. Throughout the experiments of this work, the value of $\epsilon_\text{FP}=10^{-6}$ did not yield any false positives, but to test this assumption we consider a much higher value of $\epsilon_\text{FP}=10^{-1}$ that will allow for an experimental evaluation of the resulting False Positive Rate (FPR). Two different types of false positives are considered: accusing an innocent participant after a real model leak (\textit{wrong data-owner}), and reaching an accusation when querying a non-stolen model (\textit{wrong model}). 

For the first case, all BlackCATT and BlackCATT+FR model copies trained in this work were considered, taking 100 random collusions for experiments with 20 data-owners and 500 for those with 40 and 60 data-owners. The resulting FPRs are reported in  Table~\ref{tab:exp_fpr},\footnote{If multiple data-owners surpassed the accusation threshold at the same time, the one with the highest suspicious score was considered.} and fall below the permitted $\epsilon_\text{FP}$, even with a greater number of data-owners. 

For the second case, 250 independently trained, non-watermarked models\footnote{Matching the network architecture to maximize the potential transferability of the learned features~\cite{MoosaviDezfooli16}, randomly initialized and trained for 100 epochs on the full dataset, otherwise with the same training configuration used throughout this work.} were queried as suspicious leaks in the BlackCATT+FR ResNet18 on CIFAR-100 with 60 data-owners. As shown in Table~\ref{tab:exp_fpr}, the experimental FPR of these accusation attempts approximates the theoretical bound, suggesting that while label-specific biases may exist in the trigger set, the transferability of the unique watermarks to unrelated models is still limited. 

Overall, the hypothesis of triggers being independent from labels cannot be expected to be true in this case, but experimental results show that it can still be a reasonable approximation for the number of data-owners considered in this work.

\begin{table}[ht]
\centering
\caption{Experimental FPR for an auxiliary $\epsilon_\text{FP}=10^{-1}$}\label{tab:exp_fpr}
\begin{tabular}{r|ccc|}
\cline{2-4}
\multicolumn{1}{l|}{}& \multicolumn{3}{c|}{Number of colluders $c$}\\ 
\cline{2-4}
\multicolumn{1}{l| }{} & \multicolumn{1}{c|}{1} & \multicolumn{1}{c|}{2} & 5\\ \hline
 \multicolumn{1}{|r|}{Wrong data-owner (N=20) }& \multicolumn{1}{c|}{\textbf{0.000}}& \multicolumn{1}{c|}{\textbf{0.000}}& 0.010\\ \hline
  \multicolumn{1}{|r|}{Wrong data-owner (N=40) }& \multicolumn{1}{c|}{\textbf{0.00}}& \multicolumn{1}{c|}{\textbf{0.000}}& 0.042\\ \hline
   \multicolumn{1}{|r|}{Wrong data-owner (N=60) }& \multicolumn{1}{c|}{\textbf{0.00}}& \multicolumn{1}{c|}{\textbf{0.000}}& 0.010\\ \hline
 \multicolumn{1}{|r|}{Wrong model (N=60)}& \multicolumn{1}{c|}{0.084}& \multicolumn{1}{c|}{-}&-\\ \hline

\end{tabular}
\end{table}
\section{Related Works}\label{sec:relworks}
The closest work related to the proposed BlackCATT are~\cite{RodriguezLois24} or its extension in~\cite{RodriguezLois25}, which consider the same Tardos-based accusation philosophy and the use of task arithmetic not to erase the unique watermark with each iteration. However,~\cite{RodriguezLois24,RodriguezLois25} do not provide any distinct embedding approach,\footnote{The work in ~\cite{RodriguezLois24} only proposes the use of drop-out regularization, and ~\cite{RodriguezLois25} studies the effect of exponentially weighting the most relevant parameters in the watermarking task, neither approaches have a significant effect in the networks and tasks considered in this work.} and for highly over-parameterized or deep, non-linear models, this naive embedding results in learned triggers that are vulnerable to collusion attacks, rendering the Tardos-based accusation useless, as shown with the Vanilla approach in the experimental results. BlackCATT, as presented in this paper, offers a practical embedding technique through a novel collusion-aware loss term and the progressive optimization of the shared trigger set. Additionally, some watermarking works have already introduced Continual Learning concepts in their embedding approach~\cite{Shao23} but
BlackCATT+FR is, to the best of our knowledge, the first attempt to tackle the challenge of parameter drift in numerically different model copies through functional regularization. 

As previously mentioned, other works consider unique black-box watermarks without addressing the threat of collusion~\cite{Fang-Qi_Li21,Xu25}. While the method in~\cite{Fang-Qi_Li21} motivates the Vanilla approach used in Section~\ref{sec:exp},~\cite{Xu25} was not included in the comparison because it is incompatible with the current threat model: the unique watermark is embedded into a dedicated subset of parameters, separate from those responsible for the main task and chosen after the main task has sufficiently converged, which is inherently vulnerable to collusion and earlier round theft. 

The current work focuses on traitor tracing, but watermarking in FL has also been explored for a variety of other purposes, such as aggregator-side watermarking to detect Byzantine attacks~\cite{Zheng22}, data-owner-side watermarking to prove their individual contribution~\cite{Bowen_Li22,Zhang24,Luo25}, or to verify the unlearning of their private data~\cite{XGao25}. Some studies also address a more realistic scenario with an untrusted aggregator, and propose schemes compatible with model encryption, either through trusted clients~\cite{Yang22} or watermarking embedding under homomorphic encryption~\cite{Lansari24}.

In the context of traitor tracing, a different approach has been proposed to design anti-collusion transformations~\cite{Cheng24} that degrade the performance of the merged model. Unfortunately, this is not compatible with the collaborative nature of FL, as it would not allow data-owners to share productive task updates among model copies.

Several works have also explored adversarial optimization of triggers from different perspectives, from designing more efficient trigger sets prior to the embedding~\cite{Fang-Qi_Li22}, to using the adversarial triggers themselves as proof, either as a fingerprint of the specific instance of the trained model~\cite{Peng22}, or deliberate examples left uncorrected during adversarial learning~\cite{Thakkar24}.

Additionally, while this work considers a pre-existing auxiliary dataset for the functional regularization at the aggregator, other approaches have proposed generating representative samples through data-owner-trained generative networks~\cite{Fang-Qi_Li21} or data-free distillation~\cite{Fang-Qi_Li22}.

Finally, the research on attacks and defenses of black-box watermarking is very active, with many efforts on both sides: developing improved techniques on detecting~\cite{XLiu23} and removing~\cite{Wang19,Chen22,Lu23} backdoors from trained models, and also designing schemes that will survive such attacks~\cite{Fang-Qi_Li21_2,Gao25,Zhao22}. Considering the limited scope of this work, BlackCATT is not meant to be a complete defense against all possible attacks. Instead, it provides a framework aimed at simple collusion attacks, which are particularly destructive to traitor-tracing approaches in prior work. 

\section{Conclusions and Future Work} \label{sec:concl}
To the best of our knowledge, BlackCATT is the first practical black-box traitor tracing scheme for FL. It provides a general, collusion-resistant embedding method that quickly and efficiently watermarks model copies through the simultaneous optimization of a shared trigger set, and when necessary, mitigates model drift via functional regularization to preserve main task performance. Experimental results demonstrate that BlackCATT is effective across different architectures and datasets, scales beyond small groups of data-owners, and remains robust to both collusion attacks and further processing of leaked models. Additionally, the performance of the scheme can be further improved if the aggregator is allowed additional computational resources, particularly through a larger shared trigger set or an increased number of optimization iterations after each training round.

While these results open the door to practical black-box traitor tracing in FL, many real-world challenges fall out of the scope of this work. Firstly, all experimental results were obtained assuming i.i.d. datasets across data-owners, so extending this scheme to non-i.i.d. settings may present additional difficulties. Secondly, the assumption of a fully trusted aggregator is restrictive, as the security of the scheme relies on the incorruptibility of this participant, who could also be incentivized to collude with malicious participants. Additionally, although experiments show that it is possible to attempt a stealthier trigger set, it does not achieve the same performance or effectiveness as the default artificial one, whose distinct distribution likely contributes to the effectiveness of functional regularization. Moreover, the proposed method has only been evaluated on classification tasks. Extending the approach to other architectures, such as generative models, would require auxiliary mechanisms, for example training a private classification head alongside model copies to map query outputs to a one-dimensional representation. Finally, the adversarial training of the triggers could potentially compromise the independence assumptions in theoretical Tardos analysis. These limitations highlight clear directions for future work, which can build upon BlackCATT to design more realistic and robust solutions.

\bibliographystyle{ieeetr}
\bibliography{bibliography}

%

\begin{IEEEbiography}[{\includegraphics[width=1in,height=1.25in,clip,keepaspectratio]{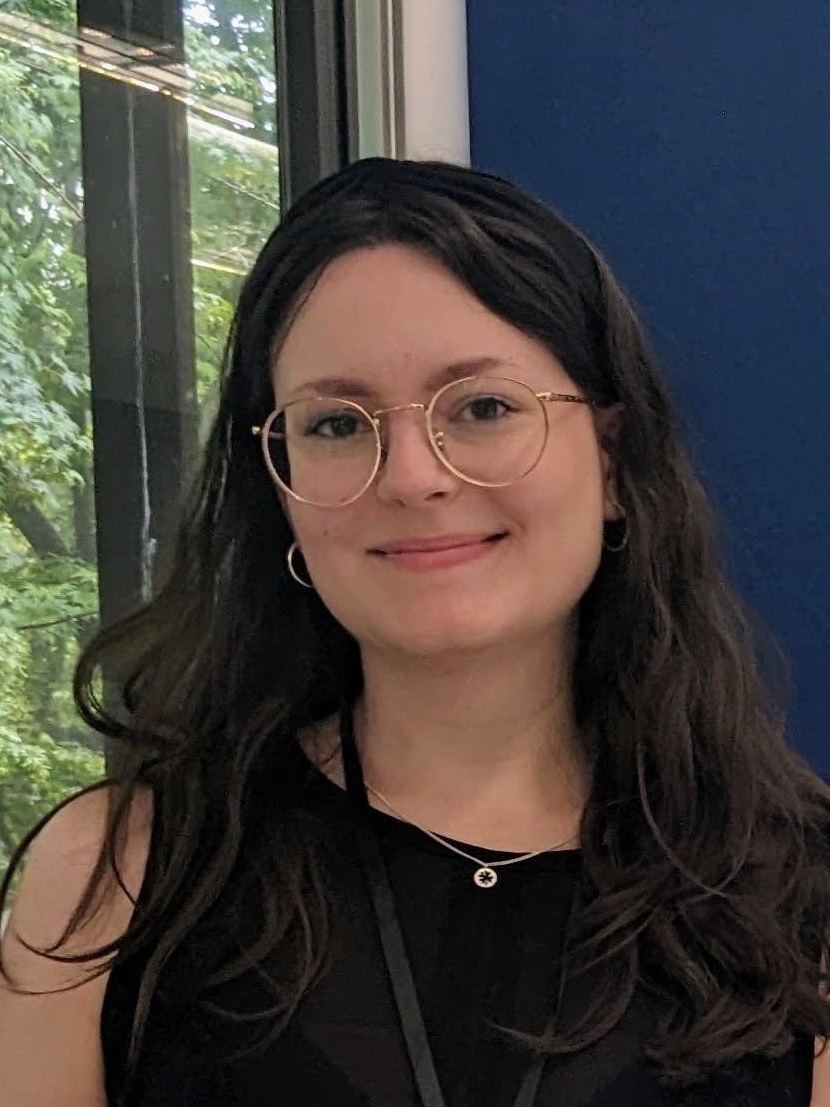}}]{Elena Rodríguez-Lois} received the B.Sc. degree (Extraordinary Graduation Award) in Industrial Electronics and Automation Engineering from the University of Vigo, Spain, in 2019, and the M.Sc. degree in Telecommunication Engineering from the same university in 2022. She is currently pursuing the Ph.D. degree at the University of Vigo under the Spanish FPU fellowship. Her research focuses on multimedia security, with a particular emphasis on watermarking deep neural networks in federated learning. She also participated in a European H2020 research project on steganalysis and image forensics.

\end{IEEEbiography}

\begin{IEEEbiography}[{\includegraphics[width=1in,height=1.25in,clip,keepaspectratio]{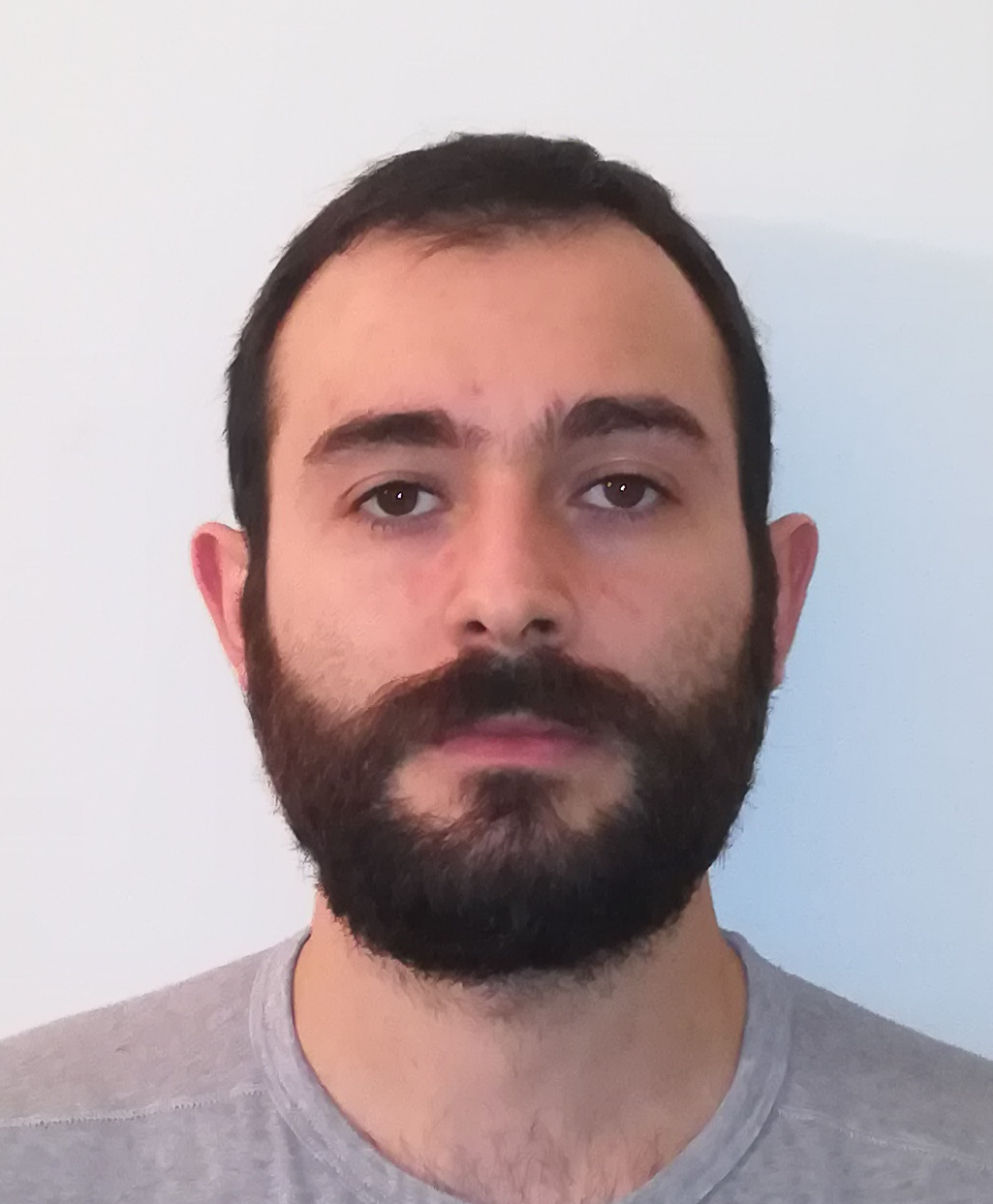}}]{Fabio Brau} holds a Master’s degree in Mathematics (Pisa, 2019) and a Ph.D. in Computer Engineering (Pisa, 2023), where he focused on the certification of robustness for deep neural networks. His current work spans Adversarial Machine Learning and Trustworthiness of AI Systems, for CV and NLP systems. He has also contributed to the theoretical and practical study of Lipschitz-constrained architectures for enhancing model stability and certifiable robustness. He published in top-tier AI/ML conferences such as NeurIPS, CVPR, AAAI and, in journals such as IEEE TPAMI. He is currently involved in European research projects on trustworthy AI, and serves as a reviewer for several conferences and journals.
\end{IEEEbiography}

\begin{IEEEbiography}[{\includegraphics[width=1in,height=1.25in,clip,keepaspectratio]{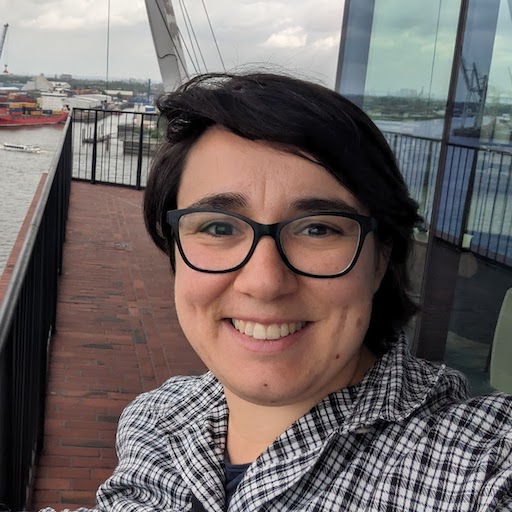}}]{Maura Pintor} is an Assistant Professor at the University of Cagliari, Italy. She received her PhD in Electronic and Computer Engineering (with honors) in 2022 from the University of Cagliari. Her research interests include trustworthy robustness evaluation of ML models, with a focus on optimizing and debugging adversarial attacks.
She is AC for NeurIPS, CAE for IEEE TIFS, and reviewer for IEEE S\&P and CVPR, and for several top-tier journals (IEEE TIFS, IEEE TDSC, IEEE-TNNLS, IEEE TIP, ACM TOPS). 
She is member of IEEE, ACM, IAPR, and Ellis. She is also a member of the Information Forensics and Security Technical Committee.
\end{IEEEbiography}

\begin{IEEEbiography}[{\includegraphics[width=1in,height=1.25in,clip,keepaspectratio]{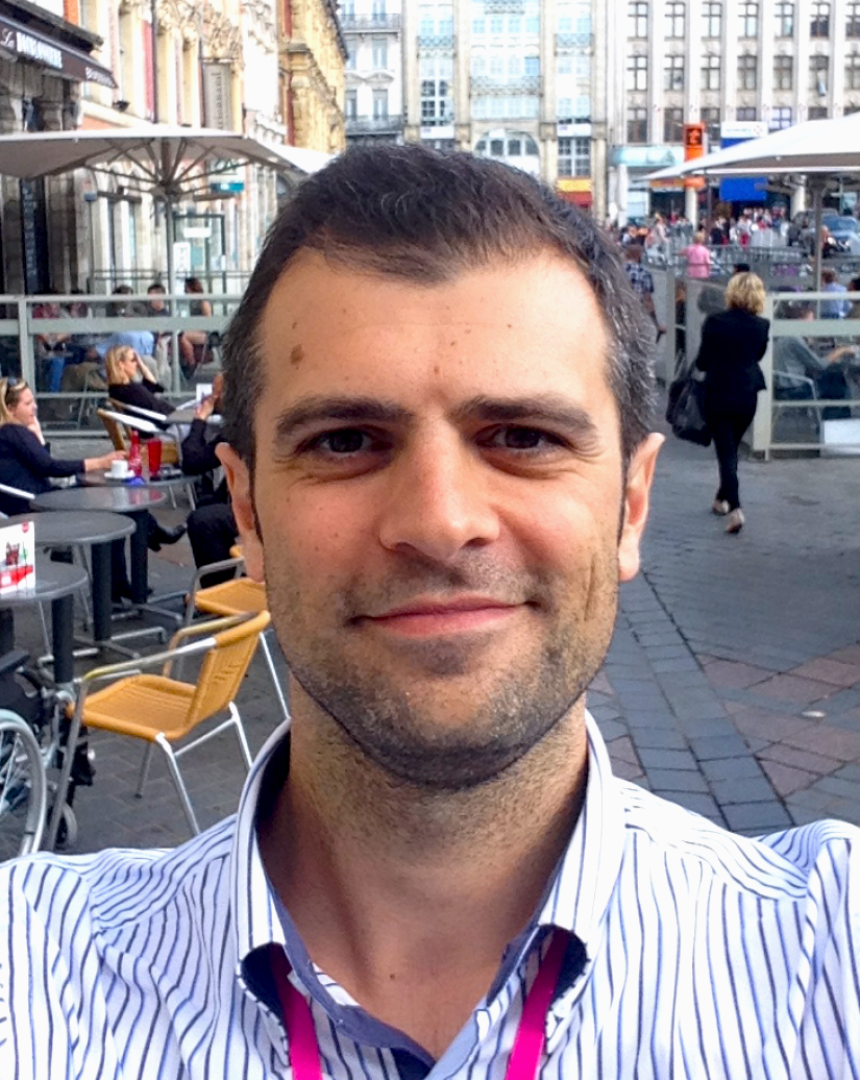}}]{Battista Biggio} (MSc 2006, PhD 2010) is Full Professor at the University of Cagliari, Italy. He has provided pioneering contributions in machine learning security, playing a leading role in this field. His seminal paper on ``Poisoning Attacks against Support Vector Machines'' won the prestigious 2022 ICML Test of Time Award. His work on ``Wild Patterns'' won the 2021 Best Paper Award and Pattern Recognition Medal from Elsevier Pattern Recognition. He has managed more than 10 research projects, and serves as a PC member of ICML and USENIX Security, and as Area Chair of NeurIPS. He chaired IAPR TC1 (2016-2020), and served as Associate Editor for IEEE TNNLS, IEEE CIM, and Elsevier PRJ. He is now Associate Editor-in-Chief for PRJ. He is also Fellow of IEEE, Senior Member of ACM, and member of IAPR and ELLIS. 
\end{IEEEbiography}

\begin{IEEEbiography}[{\includegraphics[width=1in,height=1.25in,clip,keepaspectratio]{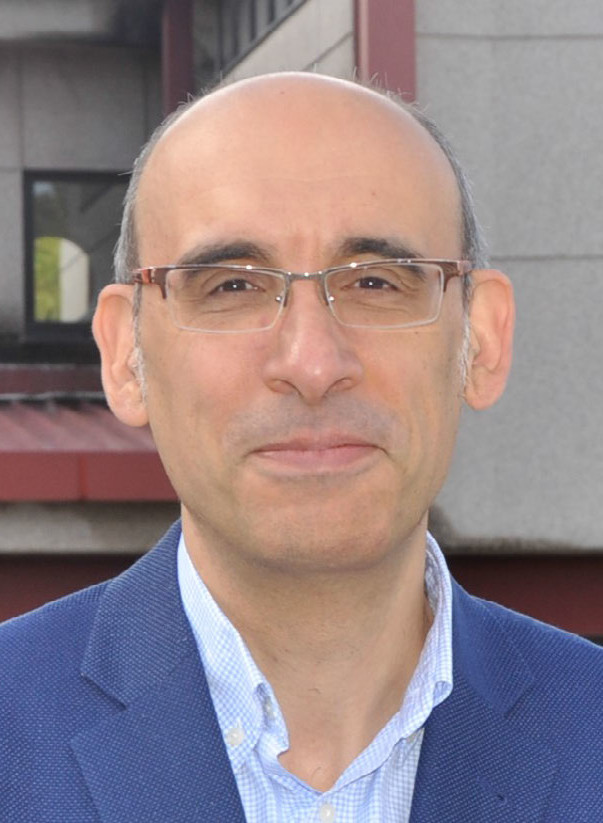}}]{Fernando P\'erez-Gonz\'alez} (Fellow, IEEE) received the degree in telecommunication engineering from the University of Santiago, Santiago, Spain, in 1990, and the Ph.D. degree in telecommunications engineering from the University of Vigo, Vigo, Spain, in 1993.

He is currently a Professor with the School of Telecommunication Engineering, University of Vigo. From 2007 to 2010, he was the Program Manager of the Spanish National Research and Development Plan on Electronic and Communication Technologies, Ministry of Science and Innovation. From 2009 to 2011, he was the Prince of the Asturias Endowed Chair of Information Science and Technology, The University of New Mexico, Albuquerque, NM, USA. From 2007 to 2014, he was the Executive Director of the Galician Research and Development Center in Advanced Telecommunications. He has been a Principal Investigator with the Signal Processing in Communications Group, University of Vigo, which participated in several European projects. He has coauthored over 70 papers in leading international journals, 180 peer-reviewed conference papers, and several international patents. His research interests include the areas of digital communications, adaptive algorithms, privacy enhancing technologies, and information forensics and security.

Dr. P\'erez-Gonz\'alez has served as Associate Editor of IEEE Signal Processing Letters (2005–2009), IEEE Transactions on Information Forensics and Security (IEEE-TIFS, 2006–2010, 2023–present), and was Editor-in-Chief of the EURASIP International Journal on Information Security (2017–2022). From 2019 to 2021, he was Senior Area Editor for IEEE-TIFS. 
\end{IEEEbiography}







\end{document}